\documentclass[conference]{IEEEtran}
\usepackage{cite}
\usepackage{amsmath,amssymb,amsfonts}
\usepackage{algorithmic}
\usepackage{graphicx}
\usepackage{textcomp}
\usepackage{xcolor}
\def\BibTeX{{\rm B\kern-.05em{\sc i\kern-.025em b}\kern-.08em
    T\kern-.1667em\lower.7ex\hbox{E}\kern-.125emX}}

\begin{document}

\title{Multitask Learning for Polyphonic Piano Transcription, a Case Study}

\author{\IEEEauthorblockN{Rainer Kelz}
\IEEEauthorblockA{\textit{Austrian Research Institute} \\
\textit{ for Artificial Intelligence (OFAI)}\\
Vienna, Austria \\
rainer.kelz@ofai.at}
\and
\IEEEauthorblockN{Sebastian B{\"o}ck}
\IEEEauthorblockA{\textit{Austrian Research Institute} \\
\textit{ for Artificial Intelligence (OFAI)}\\
Vienna, Austria \\
sebastian.boeck@ofai.at}
\and
\IEEEauthorblockN{Gerhard Widmer}
\IEEEauthorblockA{\textit{Institute of Computational Perception} \\
\textit{Johannes Kepler University}\\
Linz, Austria \\
gerhard.widmer@jku.at}
}

\maketitle

\begin{abstract}
Viewing polyphonic piano transcription as a multitask learning problem,
where we need to simultaneously predict onsets, intermediate frames and
offsets of notes, we investigate the performance impact of additional
prediction targets, using a variety of suitable convolutional
neural network architectures. We quantify performance differences
of additional objectives on the large MAESTRO dataset.
\end{abstract}


\section{Introduction}\label{sec:introduction}
Instrument specific polyphonic transcription refers to the process of transforming an audio signal containing multiple, concurrently sounding notes of a single instrument into a symbolic form. In machine learning terms, this has characteristics of both multilabel problems and structured prediction problems. Polyphonic transcription can be seen as a multilabel problem, because multiple notes sound concurrently, and hence for a short snippet of audio, we need to infer a vector that indicates which notes are active in this snippet. It can also be seen as a structured prediction problem, because ultimately we would like to obtain sets of discrete structures, denoting musical notes in a symbolic score.

Polyphonic transcription can be posed as a multitask learning problem as well, by deriving multiple prediction targets from a single groundtruth. Combining the multitask learning paradigm with deep representation learning, we hope to obtain layered, shared representations that are useful for a range of tasks. The shared representation might also decrease learning difficulty for tasks which are hard to learn on their own. In a scenario with two loosely related tasks, where one task has only very sparse label information compared to the other one, the sparsely labeled task might benefit by using (parts of) the representation learned for the densely labeled task. To pick a concrete example related to polyphonic piano transcription: it is difficult to train networks that predict pitched onsets or offsets in isolation, as the labels are very sparsely distributed in time. These tasks become easier to learn when the network needs to solve additional tasks, such as predicting all intermediate frames of a note, as this learning signal is comparatively much more densely distributed in time.

We would like to note that multitask learning is related to transfer learning in the sense that in both cases we seek to exploit shared representations across tasks. The difference lies in how the shared representations are obtained. In multitask learning, the representation is obtained simultaneously for all tasks, whereas for transfer learning, tasks are learned sequentially and therefore shared representations are obtained sequentially as well. In many cases, the performance on the tasks used to bootstrap the representation is of little concern, we are usually only interested in the performance on the last task that uses (parts of) the transferred representation.

Formally, in multitask learning \cite{caruana_2012}, we generally have multiple
\begin{align*}
\text{prediction targets}:\, & \{\mathbf{y}^{(m)} \}_{m=1}^M \\
\text{predictions}:\,          & \{\hat{\mathbf{y}}^{(m)}\}_{m=1}^M \\
\text{loss functions}:\,       & \{L^{(m)}(\mathbf{y}^{(m)}, \hat{\mathbf{y}}^{(m)})\}_{m=1}^M
\end{align*}

where $M$ is the number of tasks.

To be able to minimize these losses jointly, an aggregate function of the individual loss functions is used. The typical example would be a weighted sum of losses:
\[
L = \sum_{m=1}^M \lambda^{(m)} \cdot L^{(m)}
\]
The questions we would like to address in this case-study are to which extent do additional learning targets influence single-task performance, are these targets actually learnable from the available data, and if so, how do they affect the optimization difficulty, all in the context of obtaining a (low-level) polyphonic piano transcription system. As our experimental results will show later on in \mbox{Section \ref{sec:results}}, it is indeed beneficial to include surrogate tasks when doing multitask learning, even if the surrogate targets are themselves difficult to predict. Given the right choice of network architecture this can be done in a convenient, stable and computationally efficient way.

These questions can be answered without considering the additional complexity of the structured prediction aspect of the polyphonic transcription problem, and we therefore limit ourselves to using convolutional neural networks which output multiple indicator vectors, and simply measure framewise performance, without any smoothing or aggregation along the temporal axis. Not only does this cut down on training time, it also provides low-complexity baselines for the MAESTRO dataset \cite{maestro_2018}.

\section{Related Work}\label{sec:related_work}
\begin{figure}[t]
\centerline{\includegraphics[width=0.9\columnwidth]{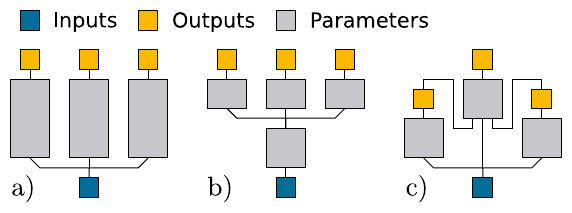}}
\caption{Sketch outlining different parameter sharing strategies for multitask networks. \textbf{a)} Multitask network architecture with no parameter sharing. \textbf{b)} Multitask network architecture with hard parameter sharing. \textbf{c)} Multitask network architecture with no parameter sharing, and conditioning on predictions from other tasks.}
\label{fig:related_work}
\end{figure}

For a comprehensive survey of multitask learning in the computer vision domain, see \cite{ruder_2017}. In the music information retrieval domain, the multitask learning paradigm has been explored in \cite{kim_2018} to investigate to which extent loosely related and unrelated MIR tasks benefit from shared representations. \cite{choi_2017} use a transfer learning approach to tackle multiple related MIR classification tasks. \cite{bittner_2018} use a multitask learning approach to obtain $f_0$-saliency maps for different instruments in different registers. \cite{chen_2018} employ multitask learning techniques for functional harmony recognition on music in symbolic form. The work of \cite{hawthorne_2018} uses two strongly related tasks, namely prediction of note onsets and intermediate note frames to push the state of the art for polyphonic piano transcription, where intermediate note frames denote all frames between note onset and offset. Two separate convolutional recurrent networks are trained, one to predict only note onsets, the other to predict intermediate note frames. The latter is conditioned on the onset predictions from the first network as an additional input, but there is no shared representation and there are no shared parameters for the two tasks. The approaches in \cite{cong_2018, lbd_2018} introduce an additional note offset prediction task, but differ in their parameter sharing strategy. No parameters are shared in \cite{cong_2018} (\mbox{Figure \ref{fig:related_work}a}), and three separate networks are trained to predict onsets, intermediate note frames and offsets respectively. In contrast, \cite{lbd_2018} use hard parameter sharing (\mbox{Figure \ref{fig:related_work}b}), where almost all network capacity is shared among the three prediction tasks, with little room for individual specialization. The models in \cite{maestro_2018, hawthorne_2018} have an additional output that predicts key strike velocity. In \cite{hawthorne_2018}, a note offset prediction network was incorporated as well, in the same fashion as the note onset prediction network, feeding its predictions into a final prediction ``head'' (\mbox{Figure \ref{fig:related_work}c}).

\section{Prediction Targets}\label{sec:prediction_targets}
\begin{figure}[t]
\centerline{\includegraphics[width=0.9\columnwidth]{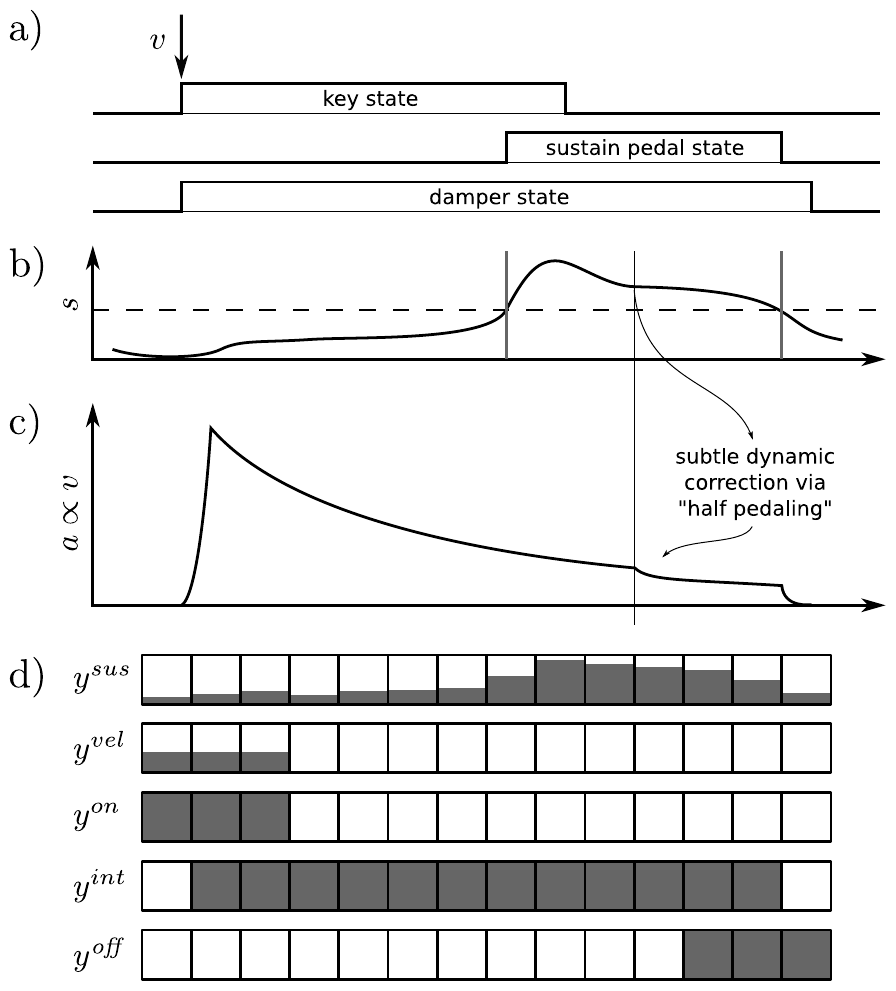}}
\caption{A sketch detailing the available information we can extract from the groundtruth. The x-axis denotes the time $t$. \textbf{a)} depicts schematically: a key being struck with a velocity $v$, subsequently being held down and finally released. While the key is held down, the sustain pedal is actuated, and the string dampers are prevented from damping until the sustain pedal is released again. \textbf{b)} shows the extent to which the sustain pedal $s$ is depressed. The dashed line indicates the threshold at which the sustain pedal is considered in full effect, and where offset correction is applied. \textbf{c)} depicts the amplitude envelope of the resulting sound, proportional to the strike velocity $v$, and modified by the sustain pedal. \textbf{d)} shows the (time quantized) prediction targets which we can derive from \textbf{a-c}.}
\label{fig:annotations}
\end{figure}

To use supervised learning techniques to train a multitask, multilabel classifier for polyphonic piano transcription, we need some form of groundtruth. The transcription groundtruth for a piano is usually given as a sequence of MIDI events that is temporally aligned with the audio. The events are parsed, and \textit{note\,on} and \textit{note\,off} events are transformed into a set of tuples $(key, onset, \mathit{offset}, velocity)$ which describe which key was pressed at what time, when it was released again, along with the velocity that the key was struck with. The key strike velocity influences mainly the volume of a note, but also modifies timbral characteristics. In the following, the extracted tuples will also be referred to as \textit{notes}.

Additional information might be available about the state of the piano pedals, e.g. if the groundtruth was obtained by means of a reproducing piano. Although there are three pedals on most modern pianos, the MAESTRO dataset \cite{maestro_2018}, produced utilizing a Disklavier, includes only information about the sustain pedal. This is arguably the most important pedal, as holding it down modifies all offsets of currently sounding notes. It does so by preventing all string dampers from touching the strings until the sustain pedal is released. This is a somewhat simplistic view on the effects the sustain pedal has on the produced sound. Mechanical pianos, as well as the Disklavier that was used to produce the dataset, enable a pianist to make subtle modifications to the sound through careful modulation of the pedal. This technique is also called ``half-pedaling''. We will see later on in \mbox{Section \ref{sec:experimental_setup}} whether the models under consideration can learn to infer the pedal state from these subtleties in the recorded audio.

All the information that the groundtruth provides is presented in \mbox{Figure \ref{fig:annotations}}. The relationship between key state (pressed, released), sustain pedal state (pressed, released) and their effect on the string damper state (up, down) is depicted in \mbox{Figure \ref{fig:annotations}a}. For the purpose of correcting the offset given by the key state to be in line with the actual offset of the note (the time when the dampers actually come to rest on the strings again, and the sound produced by the strings is muted), the control values of the sustain pedal need to be binarized. The specified range of MIDI control change messages for sustain is $s \in [0, 127]$, and we consider the sustain pedal in full effect once its value rises above $64$, which is in line with \cite{maestro_2018} and \cite{hawthorne_2018}. The range of the sustain pedal goes beyond binary values however, as outlined in \mbox{Figure \ref{fig:annotations}b}. \mbox{Figure \ref{fig:annotations}c} shows a hypothetical amplitude envelope of the sound caused by the string of the key which was struck, subtly changed by the sustain pedal action.

Finally, in \mbox{Figure \ref{fig:annotations}d} we see the time quantized prediction targets derived from the MIDI groundtruth. We have $5$ targets, all of them in the range $[0, 1]$. Targets for onsets, intermediate frames and offsets $y^{on}, y^{int}, y^{\mathit{off}}$, are strictly binary, the velocity $y^{vel}$ and the sustain pedal values $y^{sus}$ are rescaled from $[0, 127]$ into $[0, 1]$. Due to the Disklavier having $88$ keys, the prediction targets extended to all keys are thus $\mathbf{y}^{on}, \mathbf{y}^{int}, \mathbf{y}^{\mathit{off}}, \mathbf{y}^{vel} \in [0, 1]^{88}$, and as there is only one sustain pedal, its target is $y^{sus} \in [0, 1]$. To ease the optimization process, we increase label density in time for the targets $\mathbf{y}^{on}, \mathbf{y}^{\mathit{off}}$ and $\mathbf{y}^{vel}$, by applying a maximum filter of length $3$ in the time direction.

\section{Model Architectures}\label{sec:model_architectures}
\begin{figure}[t]
\centerline{\includegraphics[width=0.9\columnwidth]{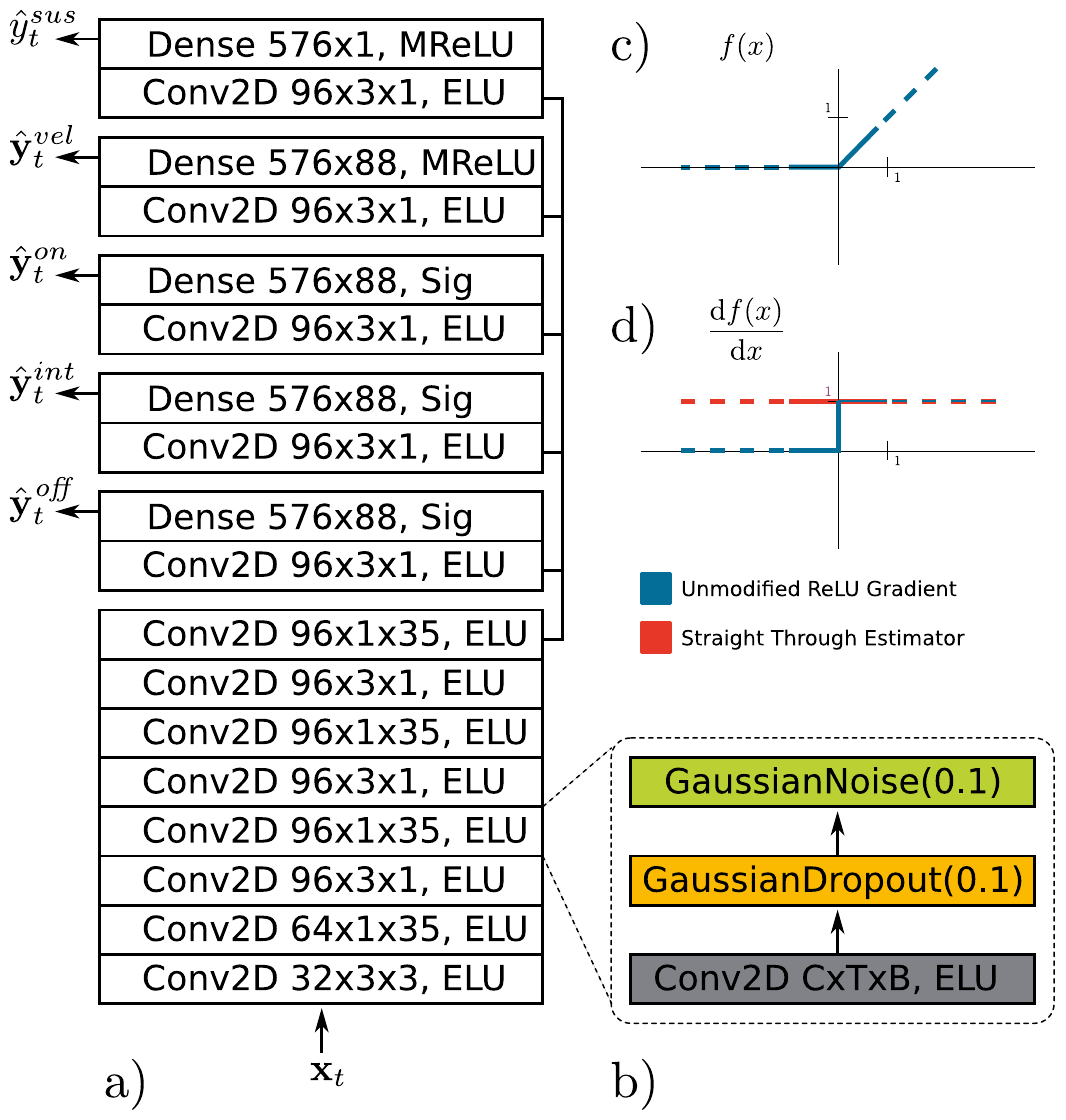}}
\caption{A sketch detailing the network architecture. Each block in \textbf{a)} labeled $\mathrm{Conv2D}\,\, \mathrm{C}\times\mathrm{T}\times\mathrm{B},\, \mathrm{ELU}$ consists of a convolutional layer with $\mathrm{C}$ feature maps, and a kernel extending $\mathrm{T}$ frames in time, and $\mathrm{B}$ bins in frequency, followed by an $\mathrm{ELU}$ nonlinearity \cite{clevert_2015} and two noise injection layers \textbf{(b)}. One introduces multiplicative gaussian noise, and one additive gaussian noise. Please see the text for an explanation of subfigures \textbf{c} and \textbf{d}.}
\label{fig:network_architecture}
\end{figure}

The model described in \cite{lbd_2018} is very economical in terms of parameters, tuned to the size of the much smaller MAPS dataset \cite{emiya_2010}. We adopt this architecture, but make a few minor adaptations, such as increasing its depth and the number of feature maps, to account for the much larger MAESTRO dataset. A schematic drawing of the modified model architecture is shown in \mbox{Figure \ref{fig:network_architecture}}. An additional network architecture is also considered, which was explicitly designed for multitask learning, and is called ``cross-stitch'' network \cite{misra_2016}.

In order to do multitask learning and potentially benefit from shared representations for different tasks, we are faced with the question which representations at which layers should be shared. Although it is true that a particular representation can have more than one parametrization (or even functional form), \textit{efficiently} learning shared representations in neural networks means at some point deciding which parameters should be shared.

The prevalent parameter sharing configuration for multitask learning is \textit{hard sharing} (of which the adapted architecture depicted in \mbox{Figure \ref{fig:network_architecture}} is an example), where all tasks directly share the same parameters \cite{caruana_1993}. This is not necessarily the best strategy, as some tasks might have the need for a specialized representation, and with all the network capacity being shared, these needs could affect other tasks negatively by claiming too much capacity for themselves. Additional negative effects may manifest if we have difficult tasks with large variance losses and noisy (potentially large scale) gradients that hamper learning. Such cases contraindicate hard sharing of all layers and call for greater isolation. Related to individual task difficulty is the choice of the task weights $\lambda^{(m)}$ in the joint objective \mbox{$L = \sum_{m} \lambda^{(m)} \cdot L^{(m)}$}. Especially for hard parameter sharing schemes, some task weights need to be downscaled considerably in order to stabilize training.

One method to answer the question which layers to share is simply enumerating all possible parameter sharing variants, train each in turn and then compare results. This brute force approach quickly becomes computationally expensive, as already for a network with $L$ layers and only $2$ tasks, we would need to train and evaluate $L - 1$ sharing options, not even taking into account different task weightings and all their possible combinations.

\begin{figure}[t]
\centerline{\includegraphics[width=0.9\columnwidth]{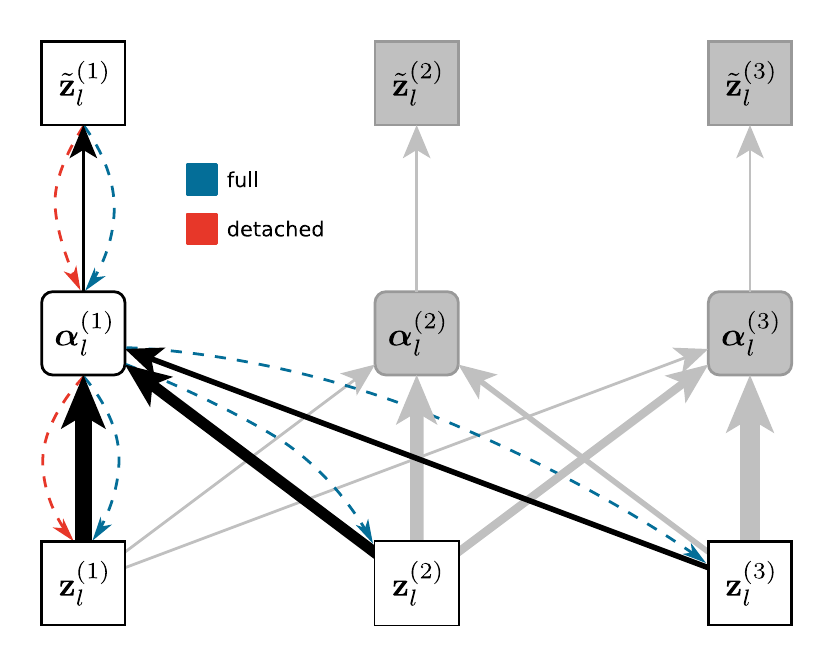}}
\caption{The two cross-stitching schemes, sketched for three task specific networks. The cross-stitch units combine all task specific representations $\mathbf{z}_{l}^{(m)}$ by linear combination with weights $\boldsymbol{\alpha}_{l}^{(m)}$ to form the input $\tilde{\mathbf{z}}_{l}^{(m)}$, which then gets passed on to the next layer. Shown in blue is the modified backwards pass for the cross-stitch units with detached feature maps. The unmodified, original backwards pass that allows gradients to flow across cross-stitch units, is shown in red.}
\label{fig:cross_stitch_schematic}
\end{figure}

Cross-stitch networks start out with a separate network for each task and try to learn at each layer which other networks' representation to draw from, via designated \textit{cross stitch units} \cite{misra_2016}. These units form a linear combination
\begin{align*}
  \tilde{\mathbf{z}}_{l}^{(m)} & = \boldsymbol{\alpha}_{l}^{(m)} \cdot \bigg[ \mathbf{z}_{l}^{(m')} \bigg]_{m'=1}^M \\
  \boldsymbol{\alpha}_{l}^{(m)} & = \bigg[ \alpha_{l,1}^{(m)}, \alpha_{l,2}^{(m)}, \dots, \alpha_{l,M}^{(m)} \bigg]
\end{align*}

of the separate representations $\mathbf{z}_{l}^{(m')}$ for the individual tasks, and pass it on to the next layer. This operation is also schematically outlined in \mbox{Figure \ref{fig:cross_stitch_schematic}}. Connecting separate, task specific representations via linear combinations frees us from having to enumerate all possible sharing configurations, and instead enables learning the amount of sharing between tasks at all layers simultaneously.
The specific cross-stitch architecture we use takes the shared part (the ``stem'') of the network depicted in \mbox{Figure \ref{fig:network_architecture}}, and simply inserts cross-stitch units after each layer. The number of feature maps for the larger layers is reduced from $96$ to $48$, to keep the total parameter count for the cross-stitch network roughly comparable to that of the hard sharing network.

We experiment with a small modification to the way gradients are backpropagated through the cross-stitch network, and test it against the unmodified version. The reason for the modification is the influence difficult tasks have on the performance of easily learned tasks. To make it precise, the attribute ``difficult'' in this context qualifies tasks that are learned much more slowly compared to the rest, and cause noisy, large scaled gradient signals that suppress more useful gradient information emanating from the easier, less noisy tasks. Cross-stitch units still allow these noisy gradients to propagate \textit{across} task-specific networks (depicted in blue in \mbox{Figure \ref{fig:cross_stitch_schematic}}), albeit to a lesser degree than if we used a hard sharing network. The backpropagation pass through the network is therefore modified to stop gradient flow across the cross-stitch units into the task-specific networks (depicted in red in \mbox{Figure \ref{fig:cross_stitch_schematic}}), preventing them from perturbing the task-specific representations. We refer to the original cross-stitch network formulation as ``full'', and our modified version as ``detached''. Gradient flow is illustrated in \mbox{Figure \ref{fig:cross_stitch_schematic}}. Note that in both cases the cross-stitch units' values $\boldsymbol{\alpha}_{l}^{(m)}$ are updated, based on $\partial L / \partial \tilde{\mathbf{z}}_{l}^{(m)}$.

All models we consider take as input a short spectrogram snippet and have the same number of outputs corresponding to the targets detailed in \mbox{Section \ref{sec:prediction_targets}}. The snippet \mbox{$\mathbf{x}_t \in \mathbb{R}^{c \times b}$}, extending $c=11$ context frames in time and $b=144$ bins in frequency, results from passing a linear STFT through a semi-logarithmic filterbank. In this filterbank, lower STFT frequencies have a linear response, higher STFT frequencies have a logarithmic response, yielding roughly $2$ bins per semitone for the logarithmic part of the filterbank. Magnitudes are mapped to a logarithmic scale as well.

The output nonlinearities producing $\hat{\mathbf{y}}^{on}, \hat{\mathbf{y}}^{int}, \hat{\mathbf{y}}^{\mathit{off}}$ are all sigmoid functions $\sigma(x) = 1 / (1 + e^{-x})$, whereas the output nonlinearities that yield $\hat{\mathbf{y}}^{vel}$ and $\hat{\mathbf{y}}^{sus}$ are (modified) ReLU units $\sigma(x) = \max(0, x)$ (shown in \mbox{Figure \ref{fig:network_architecture}c}). The reason for using modified ReLU units for these outputs is that velocity and sustain pedal are both regression targets. In order to use ReLUs as outputs, we need to address the following inconvenience first: over the course of the optimization, a ReLU unit might become inactive for all possible inputs. As soon as this happens, gradient flow is interrupted and the unit will stay inactive (zero) from then on. This is colloquially known as the ``dying ReLU problem''. We fix this issue by modifying the gradient of the ReLU (shown in blue in \mbox{Figure \ref{fig:network_architecture}d}) to behave like that for a linear function (shown in red in \mbox{Figure \ref{fig:network_architecture}d}). We borrow this idea from how ``straight through'' estimators are implemented, usually for backpropagating through stochastic neurons \cite{bengio_2013}.

\section{Experimental Setup}\label{sec:experimental_setup}
All computational experiments were conducted with the MAESTRO dataset \cite{maestro_2018}. It contains $172.3$ hours of piano performance recordings accompanied by groundtruth information in the form of precisely aligned MIDI data (alignment error is purported to be less than $3$ milliseconds). We utilize this large dataset to compare two different architecture types suitable for multitask learning: the ``hard sharing'' architecture and the ``cross-stitch'' architecture, both described in \mbox{Section \ref{sec:model_architectures}}. The multitask learning problem itself and both of these architectures require the tuning of a variety of hyperparameters. A fine grained grid search, an extensive random search, or any other automated hyperparameter optimization would overstrain our computational budget. We therefore have to partially rely on prior knowledge about which hyperparameters to adjust and in some cases fall back on heuristics that have proven to work well in practice. All reported results stem from models with roughly the same amount of parameters. Models were trained on the same training data, and evaluated on the validation split of the dataset. Training lasted for $20.000 \times 34 = 680.000$ gradient update steps in all cases. The minibatch size was fixed to $64$ samples for all models. At a frame rate of $50$ fps, the training set comprises $\approx 25.2 \cdot 10^6$ frames. This means that after training the models have seen all training data roughly \mbox{$43.5 \cdot 10^6 / 25.2 \cdot 10^6 = 1.7$ times}, to provide an estimate of how many ``epochs'' that would be.

Most hyperparameters, such as minibatch size ($64$), choice of optimizer (gradient descent with momentum (0.9) and Nesterov correction \cite{nesterov_1983}), parameter initialization (drawn from a uniform distribution according to \cite{glorot_2010}, colloquially called ``Glorot, Uniform'') as well as kernel sizes were chosen based on prior knowledge that they lead to acceptable performance for the smaller model described in \cite{lbd_2018}. The number of feature maps for the scaled up version of the model was chosen as a trade off between capacity and training time.

The learning rate $\eta$ is arguably the most important hyperparameter when training deep convolutional networks. Fortunately there is a (relatively) cheap heuristic available to find stable learning rates that still enable good progress \cite{smith_2018}, and it works as follows: once all other hyperparameters are chosen, start with a tiny learning rate (on the order of $1 \cdot 10^{-8}$), make a few thousand gradient update steps while steadily increasing the learning rate up until the loss on the training set diverges. Proceed to find the learning rate associated with the minimum of this loss curve, and decrease it by about an order of magnitude. This yields a conservative learning rate estimate that provides stable learning. We used this setting to train a full set of models. In an attempt to push the envelope, this value was then increased until we ran into severe training instability issues. We marked \mbox{Tables \ref{tab:hard_sharing_lr001} - \ref{tab:cross_stitch_lr005}} with the learning rate ($\eta$) that led to the results.

Determining learning rate stability with the above heuristic has a useful side effect, as it also provides us with guidance on how to choose the task weightings $\lambda^{(m)}$, as they are task specific learning rate modifiers. For the hard sharing architecture we could set all of $\lambda^{on}, \lambda^{int}, \lambda^{\mathit{off}}$ to $1$. Training turned unstable however, when we tried to set $\lambda^{vel} > 0.5$ or $\lambda^{sus} > 0.1$. The task weightings used to achieve a particular result are noted in the tables. We deem a particular task weight combination unstable, if it diverges even after repeated restarts of the optimization process with different random parameter initialization. Such cases are marked with a ``-'' in the result tables.

For the cross-stitch architectures, we chose the same two learning rates as for the hard sharing architecture, and noticed that it was possible to weigh \textit{all} tasks equally with this type of network. We attribute this increased stability in noisy and difficult conditions to the way cross-stitch networks learn how and what to share. We can observe this behavior for both the original and the slightly modified version of cross-stitch networks (``Full'' and ``Det'' respectively, in Tables \ref{tab:cross_stitch_lr001}, \ref{tab:cross_stitch_lr005}). Finally, the initial values for $\boldsymbol{\alpha}_{l}^{(m)}$ need to be chosen. We compare an ``Imbalanced'' initialization strategy that assigns alpha values as
\[
\alpha_{l,o}^{(m)} =
\begin{cases}
  0.9 & \text{if } o = m \\
  0.1 & \text{otherwise}
\end{cases}
\]
whereas the ``Balanced'' strategy assigns $1/M$ uniformly.

\section{Results}\label{sec:results}
\newcommand{\sumT}{\sum_t}

For onsets, intermediate frames and offsets, the performance measure we report is the framewise $F_1$ score which is computed as shown in \mbox{Eq. \ref{eq:precision}-\ref{eq:fmeasure}}, with $TP[t], FP[t]$ denoting the true and false positive counts and $FN[t]$ denotes the false negative count for the frame at time $t$. The performance measure reported for the velocity and sustain pedal regressions is the coefficient of determination, $R^2$ in \mbox{Eq. \ref{eq:r2}}, denoting the fraction of explained variance of the regression target.

\begin{align}
  P & = \frac{\sumT TP[t]}{\sumT TP[t] + FP[t]} \label{eq:precision} \\
  R & = \frac{\sumT TP[t]}{\sumT TP[t] + FN[t]} \label{eq:recall} \\
  F_1 & = \frac{2 \cdot P \cdot R}{P + R} \label{eq:fmeasure}
\end{align}

\begin{align}
  \bar{y} & = 1 / T \cdot {\textstyle \sum}_t y_t  \\
  \hat{y_t} & = f_{\theta}(\mathbf{x}_t) \\
  R^2 & = 1 - \frac{{\textstyle \sum}_t (y_t - \hat{y}_t)^2}{{\textstyle \sum}_t (y_t - \bar{y})^2} \label{eq:r2}
\end{align}

Due to restrictions on our computational budget we have only point estimates available and would like to preemptively state that comparisons between hyperparameter settings, or conclusions about apparent correlations, need to be made carefully or not at all. In the same cautionary vein, only architectures trained with the same learning rate should be compared.

In \mbox{Tables \ref{tab:cross_stitch_lr001} and \ref{tab:cross_stitch_lr005}}, which contain the results for cross-stitch networks, the abbreviations ``Det'' and ``Full'' in the column ``Type'' refer to the ``detached'' and ``full'' architectures detailed in \mbox{Section \ref{sec:model_architectures}}. Column ``Init'' in the same tables, containing abbreviations ``Bal'' and ``Imb'', refer to the ``Balanced'' and ``Imbalanced'' initialization schemes for the cross-stitch units respectively.

If we take a cursory glance at the results we might be tempted to conclude that all comparable models perform almost on par with each other. Models trained with the larger learning rate setting give better performance and have less standard deviation across different task weight configurations. Looking closer, we can see that the mean results for the onsets, intermediate frames and offset detection tasks in \mbox{Tables \ref{tab:hard_sharing_lr005} and \ref{tab:cross_stitch_lr005}} are highly similar. However, cross-stitch networks achieve the same or better performance across tasks at a \textit{much reduced} computational cost, as evidenced by the naturally higher mean performance for velocity and sustain regression, by allowing us to avoid the incremental probing for the largest possible task weighting for difficult tasks that still admits stable optimization.

Looking at the results for the hard sharing model trained with the lower learning rate, which are shown in \mbox{Table \ref{tab:hard_sharing_lr001}}, it appears that velocity regression performance and onset performance are negatively correlated. In contrast, this apparent correlation disappears if we consider \mbox{Table \ref{tab:hard_sharing_lr005}}, where onset detection performance turns out to be unaffected. For the cross-stitch networks, where there is less competition for network capacity, this is true as well.

There is a positive correlation observable between velocity and sustain regression performance, in both \mbox{Tables \ref{tab:hard_sharing_lr001} and \ref{tab:hard_sharing_lr005}}. For a particular fixed velocity task weighting, if we do not include the sustain pedal task during optimization \mbox{($\lambda^{sus} = 0$, table rows 4 and 7)}, the performance on the velocity task stays low. As soon as the sustain pedal is included \mbox{($\lambda^{sus} > 0$, table rows 5,6,8,9)}, the performance on the velocity task increases, even though its task weight $\lambda^{vel}$ stays fixed. We could not find a stable combination of task weights with both $\lambda^{vel} > 0.5$ and $\lambda^{sus} > 0$, so this potential correlation stays unobservable for larger velocity task weights in hard sharing networks.

Due to the greater stability of cross-stitch networks, which enable equal task weighting for all tasks, this is not an issue, as we can see from much better results for both velocity and sustain pedal performance in \mbox{Tables \ref{tab:cross_stitch_lr001} and \ref{tab:cross_stitch_lr005}}. A possible explanation for this behavior is that both targets are related to the volume of the input signal. Velocity is related to the volume of individual notes, and sustain is related to the volume of multiple, simultaneously sounding notes. This requires the representation to at least partially preserve pitch specific volume information from the input.


\begin{table}[t]
\caption{Hard Sharing, $\eta = 0.01$}
\begin{center}
\begin{tabular}{|p{0.01cm}|l|l||c|c|c||c|c|}
\hline
 & $\lambda^{vel}$ & $\lambda^{sus}$ & On $F_1$ & Int $F_1$ & Off $F_1$ & Vel $R^2$ & Sus $R^2$ \\
\hline
\tiny{1} & 0 & 0 & 0.8169 & 0.7992 & 0.4767 & - & - \\
\tiny{2} & 0 & 0.01 & 0.8167 & 0.7990 & 0.4815 & - & 0.0832 \\
\tiny{3} & 0 & 0.1 & 0.8159 & 0.7996 & 0.4810 & - & 0.1274 \\
\hline
\tiny{4} & 0.01 & 0 & 0.8194 & 0.8015 & 0.4857 & 0.0398 & - \\
\tiny{5} & 0.01 & 0.01 & \textbf{0.8397} & 0.8112 & 0.5355 & 0.1625 & 0.1036 \\
\tiny{6} & 0.01 & 0.1 & 0.8394 & 0.8130 & \textbf{0.5590} & 0.1774 & 0.1350 \\
\hline
\tiny{7} & 0.1 & 0 & 0.8317 & 0.8091 & 0.5181 & 0.5439 & - \\
\tiny{8} & 0.1 & 0.01 & 0.8389 & \textbf{0.8130} & 0.5419 & 0.6311 & 0.1033 \\
\tiny{9} & 0.1 & 0.1 & 0.8374 & 0.8125 & 0.5356 & 0.6253 & \textbf{0.1360} \\
\hline
\tiny{10} & 0.5 & 0 & 0.8219 & 0.8014 & 0.4876 & 0.6980 & - \\
\hline
\tiny{11} & 1 & 0 & 0.8194 & 0.8005 & 0.4837 & \textbf{0.7185} & - \\
\hline
\hline
 & & mean & 0.8270 & 0.8055 & 0.5079 & 0.4496 & 0.1148 \\
 & & std & 0.0098 & 0.0059 & 0.0290 & 0.2577 & 0.0194 \\
\hline
\end{tabular}
\label{tab:hard_sharing_lr001}
\end{center}
\end{table}

\begin{table}[t]
\caption{Hard Sharing, $\eta = 0.05$}
\begin{center}
\begin{tabular}{|p{0.01cm}|l|l||c|c|c||c|c|}
\hline
 & $\lambda^{vel}$ & $\lambda^{sus}$ & On $F_1$ & Int $F_1$ & Off $F_1$ & Vel $R^2$ & Sus $R^2$ \\
\hline
\tiny{1} & 0 & 0 & 0.8487 & \textbf{0.8194} & 0.5638 & - & - \\
\tiny{2} & 0 & 0.01 & - & - & - & - & - \\
\tiny{3} & 0 & 0.1 & - & - & - & - & - \\
\hline
\tiny{4} & 0.01 & 0 & 0.8479 & 0.8178 & 0.5615 & 0.2462 & - \\
\tiny{5} & 0.01 & 0.01 & 0.8496 & 0.8173 & 0.5653 & 0.4791 & 0.1053 \\
\tiny{6} & 0.01 & 0.1 & 0.8492 & 0.8168 & \textbf{0.5795} & 0.4752 & \textbf{0.1337} \\
\hline
\tiny{7} & 0.1 & 0 & 0.8457 & 0.8173 & 0.5567 & 0.7105 & - \\
\tiny{8} & 0.1 & 0.01 & 0.8495 & 0.8175 & 0.5716 & 0.7465 & 0.1049 \\
\tiny{9} & 0.1 & 0.1 & \textbf{0.8496} & 0.8189 & 0.5726 & 0.7490 & 0.1336 \\
\hline
\tiny{10} & 0.5 & 0 & 0.8469 & 0.8174 & 0.5508 & \textbf{0.7563} & - \\
\hline
\tiny{11} & 1 & 0 & - & - & - & - & - \\
\hline
\hline
 & & mean & 0.8484 & 0.8178 & 0.5652 & 0.5947 & 0.1194 \\
 & & std & 0.0014 & 0.0008 & 0.0086 & 0.1834 & 0.0143 \\
\hline
\end{tabular}
\label{tab:hard_sharing_lr005}
\end{center}
\end{table}

\begin{table}[t]
\caption{Cross Stitch, $\eta = 0.01$}
\begin{center}
\begin{tabular}{|p{0.01cm}|l|l||c|c|c||c|c|}
\hline
 & Type & Init & On $F_1$ & Int $F_1$ & Off $F_1$ & Vel $R^2$ & Sus $R^2$ \\
\hline
\tiny{1} & Det & Bal & 0.8271 & 0.7844 & 0.4909 & 0.7270 & 0.1455 \\
\tiny{2} & Det & Imb & \textbf{0.8321} & 0.7915 & 0.5057 & \textbf{0.7392} & 0.1603 \\
\tiny{3} & Full & Bal & 0.8257 & \textbf{0.8053} & \textbf{0.5287} & 0.7118 & \textbf{0.1750} \\
\tiny{4} & Full & Imb & 0.8265 & 0.8016 & 0.5240 & 0.7223 & 0.1741 \\
\hline
\hline
 & & mean & 0.8278 & 0.7957 & 0.5123 & 0.7251 & 0.1637 \\
 & & std & 0.0025 & 0.0082 & 0.0151 & 0.0098 & 0.0120 \\
\hline
\end{tabular}
\label{tab:cross_stitch_lr001}
\end{center}
\end{table}

\begin{table}[t]
\caption{Cross Stitch, $\eta = 0.05$}
\begin{center}
\begin{tabular}{|p{0.01cm}|l|l||c|c|c||c|c|}
\hline
 & Type & Init & On $F_1$ & Int $F_1$ & Off $F_1$ & Vel $R^2$ & Sus $R^2$ \\
\hline
\tiny{1} & Det & Bal & 0.8540 & 0.8004 & 0.5642 & 0.7867 & 0.1627 \\
\tiny{2} & Det & Imb & \textbf{0.8572} & 0.8036 & 0.5624 & \textbf{0.7957} & 0.1708 \\
\tiny{3} & Full & Bal & 0.8439 & \textbf{0.8165} & \textbf{0.5656} & 0.7588 & 0.1863 \\
\tiny{4} & Full & Imb & 0.8531 & 0.8141 & 0.5641 & 0.7732 & \textbf{0.1867} \\
\hline
\hline
 & & mean & 0.8520 & 0.8087 & 0.5641 & 0.7786 & 0.1766 \\
 & & std & 0.0049 & 0.0068 & 0.0011 & 0.0140 & 0.0103 \\
\hline
\end{tabular}
\label{tab:cross_stitch_lr005}
\end{center}
\end{table}

Finally, we would like to state that these results do not reach current state-of-the-art results for the full ``Wave2Midi2Wave'' model on the validation split \cite{maestro_2018}. Onset detection $F_1$ score for this model is $95.38$, intermediate frame detection $F_1$ score is $89.58$, and offset $F_1$ was not reported separately. The transcription part of the ``Wave2Midi2Wave'' model features very large bi-directional LSTM recurrent layers \cite{graves_2005}, providing potentially arbitrarily long context aggregation over time. The fixed context ($c = 11$ time steps) of our much smaller and simpler feedforward convolutional networks, without even an HMM sequence decoding stage as in \cite{lbd_2018}, cannot compete in this regard.

\section{Conclusion}\label{sec:conclusion}
We have explored the behavior of two different representation sharing strategies for convolutional neural networks in the context of multitask learning for polyphonic piano transcription. There is compelling experimental evidence that cross-stitch networks stabilize the learning process, make task weighting unnecessary, and lead to improved results for difficult tasks, such as key strike velocity estimation. We surmise that the models would benefit strongly from introducing recurrency or an HMM note decoding stage, but that the small network size and short training time are still the main limiting factors. Furthermore, there is evidence for a positive correlation between velocity and sustain pedal estimation performance. As expected, it proves to be extraordinarily difficult to infer the pedal state from subtle variations in the audio material, but used as a surrogate task it might still be useful.


\section{Acknowledgments}
This work is supported by the European Research Council via ERC Grant Agreement 670035, project \mbox{CON ESPRESSIONE} and the Austrian Research Promotion Agency (FFG) under the ``BASIS, Basisprogramm'' umbrella program. The Tesla K40 used for this research was donated by the NVIDIA Corporation.

\bibliographystyle{IEEEtran}
\bibliography{IEEEabrv,master}

\end{document}